\documentclass[twocolumn,letterpaper,aps,prl,
superscriptaddress,showpacs,amsmath]{revtex4}
\usepackage{graphicx}
\newcommand{\vc}[1]{\boldsymbol{#1}}

\begin{document}
\title{Kitaev-Heisenberg Model on a Honeycomb Lattice: 
Possible Exotic Phases in Iridium Oxides $A_2$IrO$_3$}

\author{Ji\v{r}\'{\i} Chaloupka}
\affiliation{Department of Condensed Matter Physics,  
Masaryk University, Kotl\'a\v{r}sk\'a 2, 61137 Brno, Czech Republic}
\affiliation{Max-Planck-Institut f\"ur Festk\"orperforschung, 
Heisenbergstrasse 1, D-70569 Stuttgart, Germany}

\author{George Jackeli}
\altaffiliation[]{Also at Andronikashvili Institute of Physics, 0177 
Tbilisi, Georgia.}
\affiliation{Max-Planck-Institut f\"ur Festk\"orperforschung,
Heisenbergstrasse 1, D-70569 Stuttgart, Germany}

\author{Giniyat Khaliullin}
\affiliation{Max-Planck-Institut f\"ur Festk\"orperforschung, 
Heisenbergstrasse 1, D-70569 Stuttgart, Germany}

\begin{abstract} 
We derive and study a spin one-half Hamiltonian on a honeycomb lattice
describing the exchange interactions between Ir$^{4+}$ ions in a family of
layered iridates $A_2$IrO$_3$ ($A$=Li,~Na).  Depending on the microscopic
parameters, the Hamiltonian interpolates between the Heisenberg and  exactly
solvable Kitaev models.  Exact diagonalization and  a complementary spin-wave
analysis reveal the presence of an extended spin-liquid phase near the Kitaev
limit and a conventional N\'eel state close to the Heisenberg limit. The two
phases are separated by an unusual stripy antiferromagnetic state, which is
the exact ground state of the model at the midpoint between two limits.   
\end{abstract}

\date{\today}

\pacs{
75.10.Jm, %Quantized spin models, including quantum spin frustration
75.25.Dk,  %Orbital, charge, and other orders, including coupling of these orders 
75.30.Et %Exchange and superexchange interactions
} 

\maketitle 

Magnetic systems exhibit, most commonly, long-range classical order at
sufficiently low temperatures.  An exception are frustrated magnets, 
in which the topology of the underlying lattice 
and/or competing interactions lead to an extensively degenerate manifold 
of classical states.  In such systems, exotic quantum phases of Mott 
insulators (spin liquids, valence bond solids, etc.) can emerge as the 
true ground states (for reviews see Refs.~\cite{Bal10,Mis05}). 
In quantum spin liquids, strong zero-point fluctuations of
correlated spins prevent them to ``freeze'' into magnetic or statically
dimerized patterns, and conventional phase transitions that break
time-reversal and lattice symmetries are avoided.  Spin liquids have attracted 
particular attention since Anderson proposed their possible connection  to
superconductivity of cuprates \cite{And87}. 

\begin{figure}[tbp]
\begin{center}
\includegraphics[width=8.2cm]{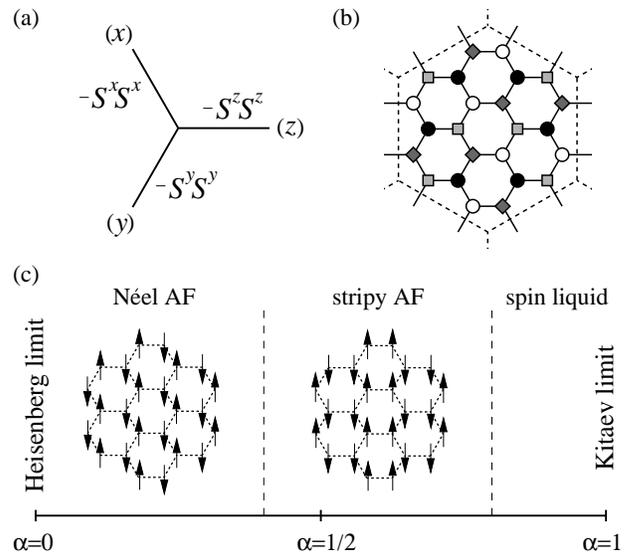}
\caption{
(a)~Three types of bonds in the honeycomb lattice and Kitaev part of the 
interaction.
(b)~The supercell of the four-sublattice system enabling the transformation of
the model (\ref{eq1}) into the Hamiltonian of a simple ferromagnet at
$\alpha=\frac{1}{2}$. This supercell with periodic boundary conditions applied
was used as a cluster for the exact diagonalization.
(c)~Schematic phase diagram: With increasing $\alpha$, 
the ground state changes from the N\'{e}el AF order to the stripy AF state
(being a fluctuation-free exact solution at $\alpha=\frac{1}{2}$) and to the
Kitaev spin liquid. See the text for the critical values of $\alpha$.
}
\label{fig:model}
\end{center}
\end{figure}

Recently, spin-liquid states of matter have been exemplified, on a
quantitative level, by an exactly solvable model by Kitaev \cite{Kit06}. His
model deals with spins one-half that live on a honeycomb lattice. The
nearest-neighbor (NN) spins interact in a simple Ising-like fashion but,
because different bonds use different spin components [see
Fig.~\ref{fig:model}(a)], the model is highly frustrated.  Its ground state is
spin-disordered and supports the emergent gapless excitations represented by
Majorana fermions \cite{Kit06}. Spin-spin correlations are, however, 
short-ranged and confined to NN pairs \cite{Bas07,Che08}. 
This may suggest the robustness
of the disordered state to spin perturbations. Indeed, Tsvelik has shown
\cite{Tsv10} that there is a window of stability for the spin-liquid state in
the Kitaev model perturbed by isotropic Heisenberg exchange. 

Finding a physical realization of this remarkable model is a great challenge, 
also because of its special properties attractive for quantum computation
\cite{Kit06}.  As the key element of the model is a bond-selective spin
anisotropy, one possible idea \cite{Jac09} is to explore Mott insulators of
late transition metal ions with orbital degeneracy, in which the bond
directional nature of electron orbitals can be translated into a desired
anisotropy of magnetic interactions through strong spin-orbit coupling. 

In this Letter, we examine the iridium oxides $A_2$IrO$_3$ from this
perspective. In these compounds, the Ir$^{4+}$ ions have an effective spin
one-half moment and form weakly coupled honeycomb-lattice planes.  Our
analysis of the underlying exchange mechanisms shows that the spin Hamiltonian
comprises two terms, ferromagnetic (FM) and antiferromagnetic (AF), in the form
of Kitaev and Heisenberg models, respectively.  The model has an interesting
phase behavior and hosts, in addition to the spin-liquid state, an unusual AF
order that is also an exact solution at a certain point in phase space.   
       
Experimental studies of iridium compounds are rather scarce, and the 
nature of their insulating behavior is not yet fully understood. In fact, 
Na$_2$IrO$_3$ was suggested as an interesting candidate for a topological band 
insulator \cite{Shi09}. Given that high temperature magnetic susceptibilities 
of Na$_2$IrO$_3$ and Li$_2$IrO$_3$ obey the Curie-Weiss law with an effective 
moment corresponding $S=1/2$ per Ir ion \cite{Fel02,Kob03,Tak10,Sin10}, 
we start here with the Mott insulator picture.  

{\it The Hamiltonian}.-- We recall that the Ir$^{4+}$ ion in the octahedral 
field has a single hole in the threefold degenerate $t_{2g}$ level hosting an
orbital angular momentum $l=1$. Strong spin-orbit coupling lifts this
degeneracy, and the resulting ground state is a Kramers doublet with total
angular momentum one-half \cite{Kim09}, referred to as ``spin'' hereafter. In
fact, it is predominantly of orbital origin, and this is what makes the
magnetic interactions highly anisotropic due to the spin-orbit entanglement of
magnetic and real spaces. In $A_2$IrO$_3$ compounds, the IrO$_6$ octahedra 
share the edges, and Ir ions can communicate through two 90$^\circ$ Ir-O-Ir
exchange paths \cite{Jac09} or via direct overlap of their orbitals.  
Collecting the possible exchange processes (discussed below) and projecting 
them onto the lowest Kramers doublet with $S=1/2$, we obtain the following 
spin Hamiltonian on a given NN $ij$ bond:  
%%%%%%%%%%%%%%%%%%%%%%%%%%%%%%%%%%%%%%%%%%%%%%%%%%%%%%%%%%%%%%%%%
\begin{equation}
{\cal H}_{ij}^{(\gamma)}=-J_{1}~S_{i}^{\gamma}S_{j}^{\gamma}+
J_{2}~{\vc S}_i\!\cdot\!{\vc S}_{j}~.
\label{eq1}
\end{equation}
%%%%%%%%%%%%%%%%%%%%%%%%%%%%%%%%%%%%%%%%%%%%%%%%%%%%%%%%%%%
Here, spin quantization axes are taken along the cubic axes of IrO$_6$
octahedra. In a honeycomb lattice formed by Ir ions, there are three distinct
types of NN bonds referred to as $\gamma(=x,y,z)$ bonds because they host the 
Ising-like $J_1$ coupling between the $\gamma$ components of spins [see
Fig.~\ref{fig:model}(a)]. The first part of Eq.~(\ref{eq1}) is thus nothing 
but the 
FM Kitaev model, and the $J_2$ term is a conventional AF Heisenberg 
model. The exchange constants $J_1$ and $J_2$ are derived from 
a  multiorbital Hubbard Hamiltonian consisting of the local interactions 
and the hopping term. The latter describes  $t_{pd\pi}$ hopping between Ir 
$5d$ and O $2p$ orbitals via the charge-transfer gap $\Delta_{pd}$, 
and a direct  $dd$ overlap $t^\prime$ between NN Ir $t_{2g}$ orbitals 
\cite{note0}. We find $J_{1}=(\eta_{1}+2\eta_{2})$ and 
$J_{2}=(\eta_{2}+\eta_{3})$. 
Hereafter, we use $4t^2/9U_d$ as our energy unit,  
where $t=t_{pd\pi}^2/\Delta_{pd}$, and $U_{d}$ stands for the Coulomb
repulsion on the same $d$ orbitals.  There are three physically distinct
virtual processes that determine the set of $\eta$ parameters and thus the 
ratio $J_2/J_1$. The $\eta_1=\frac{6J_{H}}{U_{d}-3J_H}\frac{U_d}{U_d-J_H}$ term
appears due to the multiplet structure of the excited levels induced by Hund's
coupling $J_H$ \cite{Jac09}. The processes when two holes meet at the same
oxygen site (and experience $U_p$ repulsion) and when they are cyclically
exchanged around a Ir$_2$O$_2$ square plaquette bring together
a $\eta_2=\frac{U_p}{\Delta_{pd}+U_p/2}\frac{U_d}{\Delta_{pd}}$ contribution.
Further, a direct $dd$-hopping $t^\prime$ between NN Ir $t_{2g}$ orbitals 
contributes to the Heisenberg term with exchange coupling $\eta_3=(t^\prime/t)^2$. 
It is difficult to estimate the values of all the parameters involved; 
however, we expect $\eta_1$ to be the largest, of the order of 1, 
and $\eta_{2,3}<1$.

We parametrize the exchange couplings as $J_1=2\alpha$ and $J_2=1-\alpha$
and study the properties of Kitaev-Heisenberg model (\ref{eq1}) in the whole
parameter space $0\leq\alpha\leq 1$. 

%-figure 1-----------------------------------------------------------------
%--------------------------------------------------------------------------

{\it Phase diagram}.-- At $\alpha=0$, we are left with the Heisenberg model
exhibiting the N\'eel order with a staggered moment reduced to $\langle
S^z\rangle\simeq 0.24$ \cite{Fou01}.  The opposite limit, $\alpha=1$,
corresponds to the exactly solvable Kitaev model with a short-range 
spin-liquid state \cite{Kit06}, where spin correlation functions are identically
zero beyond the NN distance and, on a given NN bond,  only the components of
spins matching the bond type are correlated \cite{Bas07}. 
 
Interestingly, the model is exactly solvable at $\alpha=\frac{1}{2}$, too. At
this point Eq.~(\ref{eq1}) reads, e.g., on a $z$-type bond, as 
${\cal H}_{ij}^{(z)}=\frac{1}{2}(S_{i}^{x}S_{j}^{x}
+S_{i}^{y}S_{j}^{y}-S_{i}^{z}S_{j}^{z})$. This anisotropic Hamiltonian can be
mapped to that of a simple Heisenberg model on all bonds simultaneously
\cite{Kha05}.  Specifically, we divide the honeycomb lattice into 
four sublattices [see Fig.~\ref{fig:model}(b)] 
and introduce the rotated operators 
$\vc{\tilde S}$: While  $\vc {\tilde S}=\vc S$ in one of the sublattices, 
$\vc{\tilde S}$ on the remaining three sublattices differs from the original 
${\vc S}$ by the sign of two appropriate components, depending on the 
sublattice they belong to. In the new basis, Eq.~(\ref{eq1}) takes the form
\begin{equation}
{\cal H}_{ij}^{(\gamma)}=
-2(2\alpha-1)~{\tilde S}_{i}^{\gamma}{\tilde S}_{j}^{\gamma}-
(1-\alpha)~{\vc {\tilde S}}_{i}\!\cdot\!{\vc {\tilde S}}_{j}~.
\label{eq3}
\end{equation}
At $\alpha=\frac{1}{2}$, the first term vanishes and we obtain the isotropic,
both in spin and real spaces, Heisenberg model 
${\cal H}_{ij}^{(\gamma)}=
-\frac{1}{2}~{\vc {\tilde S}}_{i}\!\cdot\!{\vc{\tilde S}}_{j}$ 
with FM coupling. Thus, at $\alpha=\frac{1}{2}$, i.e., at $J_1=2J_2$, the
exact ground state of model (\ref{eq1}) is a fully polarized FM state in
the rotated basis. Now consider the FM array of spins with, e.g.,
$\langle\tilde{S}^z\rangle=1/2$, 
and map it back to the original spin basis. The resulting
order corresponds to a stripy AF pattern of the original magnetic moments
depicted in Fig.~\ref{fig:model}(c). Note that such a stripy order, despite
being of AF type, is fluctuation-free at $\alpha=\frac{1}{2}$ and would thus
show a fully saturated AF order parameter.  

The above discussion suggests three possible ground state phases of the model
(\ref{eq1}) as shown in Fig.~\ref{fig:model}(c): (i) N\'eel order near
$\alpha=0$, (ii) stripy AF order around $\alpha=\frac{1}{2}$, and (iii) a
spin-liquid phase close to $\alpha=1$.

We first consider the ordered phases. Except special cases of $\alpha=0$ and
$\alpha=\frac{1}{2}$ just discussed, the Hamiltonian (\ref{eq1}) does not have
any spin-rotational symmetry.  However, a spurious $SU(2)$ continuous symmetry
and associated pseudo-Goldstone mode appear in a linear spin-wave (SW)
description. As in the case of a similar model on a cubic lattice
\cite{Kha01}, we
find that quantum fluctuations restore the underlying discrete (hexagonal)
symmetry of the model, selecting thereby the direction of ordered moments
along one of the cubic axes (of IrO$_6$ octahedra), and also open a gap in SW
spectra. Considering the quantum energy cost for rotating the order 
parameter by a small angle away from a cubic axis, we find a quantum  SW gap 
$\Delta\simeq \frac{2}{\alpha}(\alpha-\frac{1}{2})^2$ for 
$\alpha\sim\frac{1}{2}$. 

The classical phase boundary between N\'eel and stripy AF orderings is at 
$\alpha=\frac{1}{3}$, where linear SW spectra of both states develop 
zero-energy lines 
\cite{note1}, reflecting the infinite degeneracy of classical states. 
At $\alpha=\frac{1}{3}$, Eq.~(\ref{eq1}) reads, e.g., on $z$-type bonds, as
${\cal H}_{ij}^{(z)}=\frac{2}{3}(S_{i}^{x}S_{j}^{x}+S_{i}^{y}S_{j}^{y})$;
i.e., only two spin components are coupled on a given bond.  Considering
N\'eel or stripy AF with ordered spins parallel to the $z$ axis, one finds
that flipping all the spins along a zig zag chain, formed by $x$-and $y$-type
bonds, does not change classical energy. This degeneracy is again accidental
(an artifact of classical treatment) and can thus be lifted by quantum
fluctuations. They favor the N\'eel state and shift the classical phase
boundary to a larger value  $\alpha\simeq0.4$. This estimate is obtained by
comparing the energies of the N\'eel [$e_{1}\simeq -\frac{3}{16}(3-5\alpha)$]
and the stripy [$e_{2}\simeq-\frac{1}{8}(5\alpha-3+\frac{1}{\alpha})$] states
including quantum corrections via second-order perturbation theory and
matches well the numerical result found below.  

Now, we discuss the phase behavior at $\frac{1}{2}<\alpha<1$, i.e., in between
two exact solutions (stripy AF at $\alpha=\frac{1}{2}$ and a Kitaev
spin-liquid at $\alpha=1$). In terms of rotated spins, all the couplings are
of FM nature in this region [see Eq.~(\ref{eq3})]. Thus, the FM order (read
stripy AF of the original spins) is the only possible magnetic phase here to
compete with the spin-liquid state. Since the latter is stable against a weak
Heisenberg-type perturbation \cite{Tsv10}, a critical value of $\alpha$ for
the spin order/disorder transition must be located at some point less than 1.
We give its naive estimate based on the energetics of these two phases. The
energy of the stripy AF state is given above. The {\it upper} boundary
for the energy of spin-liquid state is given by the expectation value of
Eq.~(\ref{eq3}) using the exact result $\langle
S_{i}^{\gamma}S_{j}^{\gamma}\rangle=0.13$ at $\alpha=1$ \cite{Bas07}.  As a
result, we find the transition from stripy AF order to a spin liquid at
$\alpha\simeq0.86$ (close to the numerical result below). 

Single-magnon excitations fail to detect this transition (since, as said
above, there is not any other competing magnetic state). As $\alpha$
increases, the lower branch of the linear SW spectrum just gradually softens, to
become completely flat in the limit of $\alpha=1$ where the classical ground 
state is extensively degenerate \cite{Bas08}. We therefore suspect that the
instability responsible for the collapse of magnetic order resides in the
two-magnon sector \cite{note2}. Leaving this subtle issue for a 
future work, we now turn to our numerical results, which describe the 
evolution of spin correlations across the entire phase diagram. 

%-figure 2-----------------------------------------------------------------
\begin{figure}[tbp]
\begin{center}
\includegraphics[width=8.5cm]{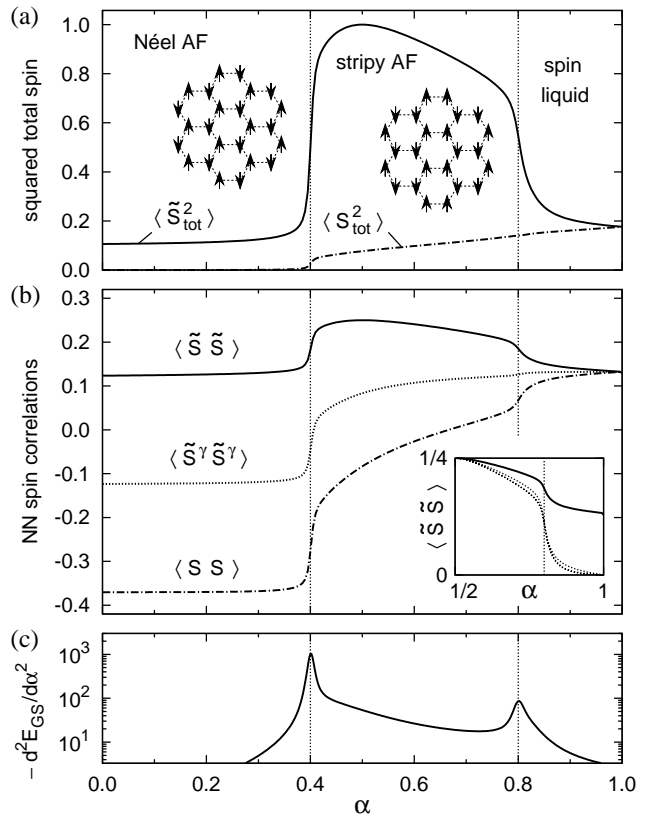}
\caption{
(a)~Squared total spin of the 24-site cluster, normalized to its value in the 
fully polarized FM state, as a function of $\alpha$. 
The solid (dot-dashed) line corresponds to the rotated (original) spin basis.
(b)~The NN spin correlations: The solid (dot-dashed) line corresponds to a scalar
product of the rotated (original) spins.  The component of the correlation
function matching the bond direction is indicated by a dotted line. This
quantity is the same in both bases.  The inset compares NN spin correlations
(solid line) above $\alpha=0.5$ with longer range spin correlations up to
third-nearest neighbors (dotted lines).
(c)~Negatively taken second derivative of the ground state energy with respect
to $\alpha$. Its maxima indicate the phase transitions at 
$\alpha\simeq 0.4$ and 0.8. 
}
\label{fig:spin}
\end{center}
\end{figure}
%--------------------------------------------------------------------------

\begin{figure}[tbp]
\begin{center}
\includegraphics[width=8.5cm]{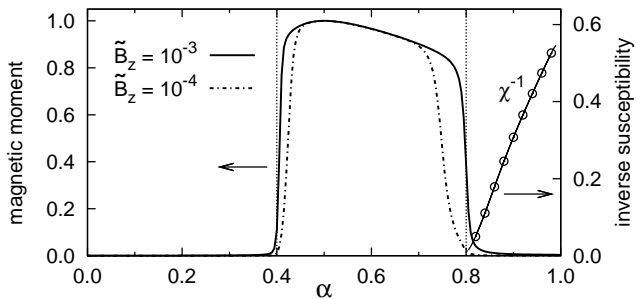}
\caption{
Magnetic moment $2\langle\tilde{S}_\mathrm{tot}^z\rangle/N$ induced by 
field $\tilde{B}_z$ (Zeeman coupled to the rotated spins). 
The circles at $\alpha>0.8$ show the inverse spin susceptibility 
to this field. 
}
\label{fig:field}
\end{center}
\end{figure}

{\it Numerical study}.-- We use the Lanczos exact diagonalization method to
study a 24-site cluster [see Fig.~\ref{fig:model}(b)] with periodic boundary
conditions. The cluster is compatible with the above discussed four-sublattice
transformation of Eq.~(\ref{eq1}) into Eq.~(\ref{eq3}). This provides an
exact reference point $\alpha=1/2$, which is useful for the interpretation of
numerical data  shown in Figs.~(\ref{fig:spin}) and (\ref{fig:field}) in terms
of the original as well as transformed spins.  

Figure \ref{fig:spin} clearly locates the two phase transitions. In particular,
a pronounced maximum in the second derivative of the ground state energy
[Fig.~\ref{fig:spin}(c)] indicates a first-order transition from 
N\'eel to stripy AF phase
at $\alpha\simeq 0.4$. The much weaker (note the {\it log} scale) and wider 
second peak at $\alpha\simeq 0.8$ suggests a second- (or a weakly first-) order 
transition from stripy AF to a spin-liquid state.

Figure \ref{fig:spin}(a) shows the squared total spins $\vc{\tilde
S}^2_{tot}$ and ${\vc S}^2_{tot}$ normalized to ${\tilde S}({\tilde S}+1)$
with ${\tilde S}=N/2$ that can be reached in the FM state.  Although these 
are not conserved quantities in the model, they characterize the phase map quite
well. In particular, a long tail of ${\vc {\tilde S}}^2_{tot}$ above 
$\alpha=0.8$ indicates a ``leakage'' of stripy AF correlations into a 
spin-liquid phase. This is also evidenced by the behavior of longer range, 
beyond NN, spin correlations that are still visible in a spin-liquid regime, 
except close to the Kitaev limit where they vanish completely [see the 
inset in Fig.~\ref{fig:spin}(b)]. 
   
Figure \ref{fig:spin}(b) highlights how the NN spin correlations evolve as 
their interactions change from one type to another. In the N\'eel state, 
where the model is more Heisenberg-like for the original spins, we 
reproduce \mbox{$\langle{\vc S}_{i}\!\cdot\!{\vc S}_{j}\rangle\simeq -0.37$}
\cite{Fou01}. At the ``hidden'' FM Heisenberg point $\alpha=1/2$, one finds 
$\langle{\vc {\tilde S}}_{i}\cdot{\vc {\tilde S}}_{j}\rangle=\frac{1}{4}$,
equally contributed by all three components of the rotated spin 
$\vc{\tilde S}$. Things change dramatically in the spin-liquid phase: Here, 
a particular component of spin correlations 
$\langle{\tilde S^{\gamma}}_{i}{\tilde S^{\gamma}}_{j}\rangle$, 
dictated by the Kitaev model, dominates. Its value of 0.132 that we find at 
$\alpha=1$ agrees well with the exact result 0.131 for an infinite
lattice \cite{Bas07}.

Finally, we discuss the response to a weak magnetic field ${\tilde B}^z$
which, in terms of original spins, linearly couples to the stripy AF order
parameter. Figure \ref{fig:field} shows that even a very weak field induces
a nearly saturated moment in the entire region of the stripy AF phase. As the
system switches to the N\'eel phase, a response to the ``stripy field''
${\tilde B}^z$ drops abruptly to zero, as expected. The induced moment
sharply reduces near $\alpha=0.8$, too, but remains finite in a spin-liquid
phase. Here the magnetization curve shows a linear dependence on ${\tilde
B}^z$, and we may extract from its slope the susceptibility
$\chi=\langle\tilde{S}_\mathrm{tot}^z\rangle/N\tilde{B}_z$. Shown in
Fig.~\ref{fig:field} is the inverse value of $\chi$ as a function of $\alpha$.
This quantity scales with 
the energy gap between the ground state and the excited states
that are accessible by the magnetic field. According to Kitaev's solution
\cite{Kit06}, these states must belong to the flux sectors located at energies
of the order of 1. The observed $\chi^{-1} \propto (\alpha-0.8)$ behavior
shows that this characteristic (spin) gap gradually softens towards the
$\alpha\simeq 0.8$ critical point, as the spin correlations beyond the 
NN distances start to grow [see Fig.~\ref{fig:spin}(b), inset].  

%-figure 3-----------------------------------------------------------------
%--------------------------------------------------------------------------

Experimental data \cite{Fel02,Kob03,Tak10,Sin10} are rather insufficient to
conclusively locate the position of  $A_2$IrO$_3$ compounds in our phase diagram. 
Also, Na/Ir site disorder \cite{Sin10} has to be kept in mind: Often, nonmagnetic
impurities induce local moments \cite{Kha97}, and this has been shown to
happen in the Kitaev model as well \cite{Wil10}.  

In conclusion, we have examined the interactions and possible magnetic states in
iridates $A_2$IrO$_3$. The obtained Kitaev-Heisenberg model shows rich
behavior including a spin liquid and unusual stripy AF phases. We
hope that these results will motivate experimental studies of layered iridates
and similar compounds of late transition metal ions, where the physics of
the Kitaev model might be within reach. 
 
We thank B. Keimer, A. Schnyder, S. Trebst, and M. Zhitomirsky 
for discussions. We are grateful to H. Takagi and A.M. Tsvelik for valuable
discussions and for communicating the unpublished results. Support from
MSM0021622410 (J.C.) and GNSF/ST09-447 (G.J.) is acknowledged. 

%%%%%%%%%%%%%%%%%%%%%%%%%%%%%%%%%%%%%%%%%%%%%%%%%%%%%%%%%%%%%%%%%


\begin{thebibliography}{99}
\bibitem{note1}
The details of SW analysis will be given elsewhere.
\end{thebibliography}
\end{document}